\documentclass[chaos,amsmath,amssymb,showpacs,groupaddress,twocolumn]{revtex4}
\usepackage{graphicx}
\usepackage{bm}
\begin{document}
\title{Topological field theory of dynamical systems. II}
\author{Igor V. Ovchinnikov}
\email{iovchinnikov@ucla.edu}\email{igor.vlad.ovchinnikov@gmail.com}
\affiliation{Department of Electrical Engineering, University of California at Los Angeles, Los Angeles, CA, 90095-1594}
\begin{abstract}
This paper is a continuation of the study [Chaos.22.033134] of the relation between the stochastic dynamical systems (DS) and the Witten-type topological field theories (TFT). Here, it is discussed that stochastic expectation values of a DS must be complemented on the TFT side by $(-1)^{\hat F}$, where $\hat F$ is the ghost number operator. The role of this inclusion is to unfold the natural path-integral representation of the TFT, \emph{i.e.}, the Witten index that equals up to a topological factor to the partition function of the stochastic noise, into the physical partition function of TFT/DS. It is also shown that on the DS side, the TFT's wavefunctions are the conditional probability densities.
\end{abstract}
\maketitle
{\bf
Approximation-free stochastic quantization procedure by Parisi and Sourlas \cite{ParisiSourlas} applied to any continuous-time (stochastic) dynamical system (DS) leads to a Witten-type topological field theory (TFT) \cite{Frenkel,Witten,TFT,Labastida}, as it was recently discussed in Ref.[\onlinecite{ChaosPaper}]. Here, the physical meanings of the Witten index and of the TFT's partition function are clarified. It is shown that it is the TFT's partition function that must be used for the evaluation of stochastic expectation values of a DS. It is also shown that the most natural interpretation of the TFT's wavefunctions is the conditional probability densities of the corresponding DS.}

\section{Introduction}
This article is the second part of Ref.[\onlinecite{ChaosPaper}], where it was shown that according the most general Parisi-Sourlas stochastic quantization procedure, \cite{ParisiSourlas} any (stochastic) continuous-time dynamical system (DS) can be looked upon as a Witten-type or cohomological TFT. \cite{TFT,Witten,Frenkel,Labastida} Previous paper was mainly concerned with the explanation of the ubiquitous emergence of the long-range correlations in various DS's known generically as complex dynamics. This paper focuses on important questions that Ref.[\onlinecite{ChaosPaper}] did not address or addressed in insufficient details.

In Sec.\ref{General}, general aspects of the stochastic quantization are presented. This section can be viewed as a version of Sec.II of Ref.[\onlinecite{ChaosPaper}], needed for the purpose of self-consistency of this paper and for the introduction of notations used later in Sec. \ref{TFTTODS}. In Secs.\ref{WittenvsPartFunction} and \ref{PhysPart}, it is shown that the physical meaning of the Witten index is the partition function of the stochastic noise up to a topological constant, and that the physical partition function of the TFT is the partition function of the DS. It is also shown that the stochastic expectation values must be defined through the physical partition function of TFT. In Sec. \ref{Goldstone}, the Goldstone theorem and the corresponding emergence of the long-range correlations are addressed. In Sec.\ref{ConditionProbabilities}, the TFT's wavefunctions are identified as the conditional probability densities of the DS. This identification establishes an interesting connection with the probability theory and provides yet another physical meaning of the phenomenon of the spontaneous topological symmetry breaking. In Sec.\ref{PhysQSym}, the physical meaning of the topological symmetry itself is revisited from the point of view of the novel interpretation of the TFT's wavefunctions. Sec. \ref{Conclusion} concludes the paper.

\section{TFT representation of DS}
\label{General}
The forthcoming discussion can be generalized to the cases of non-Gaussian and colored noises, step-like temporal evolution, infinite-dimensional phase spaces (\emph{i.e.}, higher-dimensional field theories) etc. Not to digress on unnecessary details, we will concentrate on the simplest TFT - the topological quantum mechanics (TQM) \cite{Labastida} on a finite-dimensional phase space, $M$. TQM is the result of the stochastic quantization of the following stochastic differential equation (SDE):
\begin{eqnarray}
\partial_t\varphi^i(t) + A^i(\varphi(t)) = \tilde\xi^i(t,\varphi(t)).\label{SDE}
\end{eqnarray}
Here $\varphi\in M$ are the system's variables, $A^i(\varphi)\partial/\partial\varphi^i\in TM(M)$ is the so-called flow vector field, and $\tilde\xi$'s is the stochastic Gaussian white noise:
\begin{eqnarray}
\langle\langle \tilde\xi^i(t')\tilde\xi^j(t) \rangle\rangle_\text{Ns} = \delta(t-t')g^{ij}(\varphi(t)).\label{noise}
\end{eqnarray}
The double-arrowed notation for expectation values is standard in the theory of pseudo-Hermitian models, \cite{Mostafozadeh} that the models under consideration will turn out to be. Allow us to accept this notation from the very beginning. In Eq.(\ref{noise}), $g^{ij}$ has the meaning of the noise-induced metric on $M$. By introducing veilbeins as $g^{ij}(\varphi(t)) = e^i_a(\varphi(t)) \delta^{ab} e^j_b(\varphi(t))$ we can decouple the noise from the system's variables:
\begin{eqnarray}
\tilde\xi^i(t,\varphi(t)) = e^i_a(\varphi(t)) \xi^a(t).
\end{eqnarray}
The partition function of the decoupled noise is:
\begin{eqnarray}
\mathcal{Z}_\text{Ns} = \langle\langle 1 \rangle\rangle_\text{Ns} \equiv c \int_{\xi} 1 \cdot e^{-\int_{t=0}^T (\xi^a(t))^2/2} = 1,\label{partitionfunctionofnoise}
\end{eqnarray}
where $c$ is the normalization constant that, in particular, carries the burden of the numerical consistency of integrals over infinite-dimensional spaces, such as the space of all the possible configurations of noise, $\xi^a(t)\in \mathcal{M}_\xi$, where $\mathcal{M}_\xi$ is the space of all maps from the time-circle, $t\in\mathbb{S}^1$, into the tangent space of $M$, $\xi^a(t): \mathbb{S}^1 \to TM(M) = \mathbb{R}^D$ with $D=dim M$. By thinking that the time belongs to a circle we assume periodic boundary conditions, corresponding to the case of equilibrium dynamics.

The Parisi-Sourlas method begins with the proposition to evaluate the following stochastic expectation value:
\begin{subequations}
\label{WittenOriginal}
\begin{eqnarray}
\mathcal{W} = \langle\langle [w.n.] \rangle\rangle_\text{Ns},
\end{eqnarray}
where
\begin{eqnarray}
[w.n.] &=& \int_{\varphi}\delta\left(F^i(t))\right) J. \label{windingnumber}
\end{eqnarray}
\end{subequations}
Here, yet another path-integration, besides that over the noise, is over $\varphi^i(t): \mathbb{S}^1 \to M$, $\varphi^i(t)\in \mathcal{M}_\varphi$, where $\mathcal M_\varphi$ is the space of all maps from the time circle (base space) into the phase space (target space),
\begin{eqnarray}
J = \text{det }\delta F^i(t)/\delta\varphi^j(t'),\label{Jacobian}
\end{eqnarray}
is the Jacobian, and
\begin{eqnarray}
F^i(t,\varphi,\xi) \equiv F^i(t)=  \partial_t\varphi^i+A^i-e^i_a\xi^a.
\end{eqnarray}

Using standard field theoretic methods, \cite{ChaosPaper} Eq.(\ref{WittenOriginal}) can be represented as
\begin{subequations}
\label{TQM}
\begin{eqnarray}
\mathcal{W} = \int_{\Phi} e^{\{\mathcal{Q}, \Theta\}}.\label{pathintegral0}
\end{eqnarray}
Here, $\Theta = \int_{t=0}^T (i(\partial_t\varphi^i)\bar\chi_i - \bar Q)$, is the so-called gauge fermion, defined through
\begin{eqnarray}
\bar Q(\Phi) = (- g^{ij}(iB_j - \Gamma_{jk}^l\chi^k(i\bar\chi_l))/2 -  A^i)(i\bar\chi_i),\label{Qbar}
\end{eqnarray}
\end{subequations}
with $\Gamma$'s being the Christoffel symbols, $B_i$ is an additional bosonic field called Lagrange multiplier, $\bar\chi_j$ and $\chi^i$ are the pair of additional anticommuting fields called Fadeev-Popov ghosts, and the Betty-Route-Stora-Tyutin operator of bi-graded differentiation is
\begin{eqnarray}
\{\mathcal{Q},\Phi(t)\} = (\chi^i(t), 0, B_i(t), 0),\label{BRST}
\end{eqnarray}
with $\Phi(t)\equiv(\varphi^i(t),B_i(t),\bar\chi_i(t),\chi^i(t))$ denoting the collection of all the fields.

Eq.(\ref{pathintegral0}) can be represented as:
\begin{eqnarray}
\mathcal{W} = \int_{\Phi}e^{\int_t (i(\partial_t\varphi^i)B_i + i(\partial_t\chi^i)\bar\chi_i - H(\Phi))},\label{Schrodinger}
\end{eqnarray}
where the Fokker-Planck Hamiltonian functional is given as $H = \{\mathcal{Q},\bar Q\}$, with $\bar Q$ from Eq.(\ref{Qbar}). In the Schr\"odinger picture, from Eq.(\ref{Schrodinger}) the fundamental bi-graded commutation relations, $[i\hat B_i,\hat \varphi^j]_- = [i\hat {\bar\chi}_i,\hat \chi^j]_+ = \delta^i_j$, suggests
\begin{eqnarray}
i \hat B_i = \hat \partial_{\varphi^i}\equiv \partial/\partial\varphi^i, i \hat{\bar\chi}_i = \hat \partial_{\chi^i}\equiv \partial/\partial\chi^i.
\end{eqnarray}
in the representation where $\varphi$'s and $\chi$'s are diagonal.

There is a seeming sign ambiguity in the definition of the operator $\hat{\bar\chi}_i$ because the term $i(\partial_t\chi^i)\bar\chi_i$ in the action can also be rewritten as $-i\bar\chi_i(\partial_t\chi^i)$. The sign must be chosen in such a manner that the rules of the bi-graded commutation relations of the N\"other charge of $\mathcal Q$-symmetry with the operator representations of all the fields, $\hat \Phi$, (Eq.(\ref{Commutation}) below) comply with the original action of $\mathcal Q$ in Eq.(\ref{BRST}). The sign chosen satisfies this requirement.

Wavefunctions are functions of $\varphi$'s and $\chi$'s only. Any wavefunction can be Taylor expanded in the ghosts:
\begin{eqnarray}
\psi = \sum\nolimits_{n=0}^D \psi^{(n)}, \psi^{(n)} = \psi^{(n)}_{i_1...i_n}(\varphi)\chi^{i_1}...\chi^{i_n}.
\end{eqnarray}
Due to the anticommutation composition for the ghosts, tensors $\psi^{(n)}_{i_1...i_n}$ are antisymmetric. Following Witten, \cite{FormsAndWavefunctions} we notice that the operator algebra of the theory is identical to that of the algebraic topology on $M$:
\begin{eqnarray}
\chi^i \equiv d\varphi^i\wedge, \text{ and }\hat \partial_{\chi^i} \equiv \imath_{d\varphi^i}, \label{AnaliticalTopological}
\end{eqnarray}
where $\wedge$ is the wedge product and $\imath_{d\varphi^i}$ is the interior multiplication. Therefore, the Hilbert space, $\mathcal H$, can be identified as the exterior algebra of $M$ and the wavefunctions are the forms from $\mathcal H$:
\begin{eqnarray}
\psi^{(n)} \equiv \psi^{(n)}_{i_1...i_n}(\varphi)d\varphi^{i_1}\wedge ... \wedge d\varphi^{i_n}.
\end{eqnarray}
Using the standard technique, one finds that the N\"other charge associated with the $\mathcal Q$-symmetry is $iB_i\chi^i$, which in the Schr\"odinger picture becomes the exterior derivative:
\begin{eqnarray}
\hat d \equiv \hat Q = \chi^i \hat \partial_{\varphi^i} \equiv d\varphi^i\wedge \hat \partial_{\varphi^i}.
\end{eqnarray}
Notation, $\hat Q$, is common for the TFT Literature, while $\hat d$ is the established notation in the algebraic topology. Allow us to use the notation $\hat d$, unlike in Ref.[\onlinecite{ChaosPaper}].

The equation of motion is the generalized Fokker-Planck (FP) equation:
\begin{eqnarray}
\partial_t \psi = - \hat H \psi.\label{FPEquation}
\end{eqnarray}
The FP Eq. is called here generalized in that sense that it is defined not only for the ordinary probability density, which in the TFT language is the wavefunction of the maximal ghost content, $\psi^{(D)}$, but also for the conditional probability densities that are the wavefunctions of the non-trivial ghost content, $\psi^{(n)}, n<D$. In Sec.\ref{ConditionProbabilities} we will discuss why we believe that the wavefunctions of the nontrivial ghost content must be interpreted as the condition probability densities.

The explicit form of the generalized FP Hamiltonian in Eq.(\ref{FPEquation}) can be obtained by the bi-graded symmetrization of its path-integral expression in Eq.(\ref{Schrodinger}). The easiest way to do this is to notice that the rules for the $\mathcal Q$-differentiation in Eq(\ref{BRST}) are formally the same as those for the bi-graded commutation with $\hat d$:
\begin{eqnarray}
[\hat d, \hat \Phi] = (\hat \chi^i,0,\hat B_i,0).\label{Commutation}
\end{eqnarray}
Therefore, the $Q$-exact Hamiltonian, $H(\Phi) = \{\mathcal Q, \bar Q(\Phi)\}$, becomes in the Schr\"odinger picture a $\hat d$-exact operator:
\begin{eqnarray}
\hat H = [\hat d, \hat j] = -\triangle/2-\mathcal{L}_A,\label{FPHamiltonian}
\end{eqnarray}
where $\hat j$ is the symmetrized version of $\bar Q(\Phi)$:
\begin{eqnarray}
\hat j = \hat {d}^\dagger/2 - \hat \imath_{ A},\label{current}
\end{eqnarray}
with $\hat \imath_{ A} = \hat \partial_{\chi^i} A^i$ being the interior multiplication by $A^i$ so that $\mathcal{L}_A = [\hat d, \hat \imath_{ A}]$ is Lie derivative along the flow, and with $\hat d^\dagger = -\hat \partial_{\chi^i} g^{ij}\hat \nabla_{\varphi^j}$, being the adjoint of the exterior derivative defined through the covariant derivative
$\hat \nabla_{\varphi^j} = \hat \partial_{\varphi^i} - \Gamma_{ik}^l \hat \chi^k \hat \partial_{\chi^l}$, so that $\triangle = -[\hat d, \hat d^\dagger]$ is the Weitzenb\"ock Laplacian. The expression for operator $\hat d^\dagger$ is the same as the one that follows from the operator algebra reasonings given, \emph{e.g.}, in the end of Chap. 4 of Ref.[\onlinecite{Stone}] (in this Ref.'s notations $\chi^i = \psi^\dagger\text{}^i$).

The generalized FP Hamiltonian is not Hermitian, $\hat H \ne \hat H^\dagger$. At the same time, the Hamiltonian is a real operator. This suggests that the Hamiltonian is pseudo-Hermitian \cite{Mostafozadeh} and its eigenvalues are either real or come in complex-conjugate pairs that in the DS theory are known as Ruelle-Pollicott resonances. For pseudo-Hermitian Hamitonians, the bras and kets of the eigenstates constitute a complete bi-orthogonal basis:
\begin{eqnarray}
\hat H |\alpha\rangle\rangle = \mathcal{E}_\alpha |\alpha\rangle\rangle, \langle\langle \alpha| \hat H = \langle\langle \alpha| \mathcal{E}_\alpha,\label{biorthogonal}\\
\langle\langle \alpha| \beta\rangle\rangle = \delta_{\alpha \beta}, \hat {\mathbf{1}}_{\mathcal{H}} = \sum\nolimits_{\alpha}| \alpha\rangle\rangle \langle\langle \alpha|,\label{Unity1}
\end{eqnarray}
where $\hat {\mathbf{1}}_{\mathcal{H}}$ is the unity operator on $\mathcal{H}$. In Hermitian Witten models, that correspond to the Langevin SDE's, the following notations are used
\begin{eqnarray}
|\alpha\rangle \equiv \alpha, \langle \alpha | \equiv (\star\alpha)^*, \langle \alpha| \beta \rangle = \int_M (\star\alpha)^* \wedge \beta,
\end{eqnarray}
where $\star$ is the Hodge star. In pseudo-Hermitian models, in turn, the metric on $\mathcal H$ is not trivial. To emphasize this fact, the double-arrowed notations are used for the wavefunctions. The kets are the same as in the Hermitian case $|\alpha\rangle \rangle \equiv \alpha$, while the bras include the metric on $\mathcal H$, $\eta^{-1}_{\beta\alpha} = \langle \alpha | \beta \rangle \ne \delta_{\alpha\beta}$, in the following manner:
\begin{eqnarray}
\langle \langle \alpha | \equiv \bar \alpha = \sum\nolimits_{\alpha'} \langle \alpha' |\eta_{\alpha' \alpha}, \\
\langle \langle \alpha | \beta \rangle\rangle = \int_M \bar \alpha \wedge \beta = \delta_{\beta\alpha}.
\end{eqnarray}
The metric is such that $\hat \eta \hat H \hat \eta^{-1} = \hat H^\dagger$.

All the states are divided into two major groups depending on either they do or do not break $\mathcal Q$-symmetry. States that do not break $\mathcal{Q}$-symmetry we denote as, $\theta$. These states are such that the expectation value of any $\hat d$-exact operator is zero: $\langle\langle\theta|[\hat d, \hat X]|\theta\rangle\rangle=0, \forall \hat X$, so that
\begin{eqnarray}
\langle\langle \theta| \hat d \equiv (-1)^{F_{\bar\theta}+1} (\hat d \bar \theta) = 0, \text{ and } \hat d |\theta\rangle\rangle = 0,\label{ConditionsUnbroken}
\end{eqnarray}
where $F_{\bar\theta}$ is the degree of the form $\bar\theta$. The definition of the action of the exterior derivative on a bra from the right follows from the formula of the partial integration of forms on a boundaryless, $M$:
\begin{eqnarray}
\langle\langle \alpha|\hat d|\beta\rangle\rangle =
\int_M \bar \alpha \wedge (\hat d \beta) = (-1)^{F_{\bar\alpha}+1} \int_M (\hat d \bar \alpha) \wedge \beta.\nonumber
\end{eqnarray}

The generalized Fokker-Planck Hamiltonian is $\hat d$-exact, so that all $\theta$'s have zero eigenvalues. Each cohomology class of $M$ may provide only one state that does not break $\mathcal Q$-symmetry. Indeed, if there are two states with zero eigenvalue within the same cohomology class, then their difference is an exact eigenstate, which also has zero-eigenvalue. This difference state must be a member of a bosonic-fermionic pair of eigenstates (see below) that break $\mathcal Q$-symmetry. The existence of such a bosonic-fermionic pair with zero eigenvalue is accidental however.

For the sake of completeness, allow us to discuss briefly the structure of the $\mathcal Q$-symmetric states, $\theta's$, in the deterministic limit. In this limit, bras and kets of $\theta's$ are the Poinc\^are duals of global stable (bras) and unstable (kets) manifolds of the flow. \cite{ChaosPaper,PoincareDuals} The bras and kets intersect on invariant manifolds. In case when invariant manifolds are not isolated points but are some higher-dimensional manifolds (\emph{e.g.}, the Bott-Morse case), the bras and kets of $\theta$'s must also contain factors from the cohomology of the invariant manifolds.

The majority of eigenstates, however, break $\mathcal Q$-symmetry. These states come in bosonic-fermionic pairs: $|\gamma\rangle\rangle$ and $|\gamma'\rangle\rangle = \hat d |\gamma\rangle\rangle \ne 0$. If we denote the bra of $|\gamma'\rangle\rangle$ as $\langle\langle \gamma|$, $\langle\langle \gamma|\gamma'\rangle\rangle = \langle\langle \gamma|\hat d|\gamma\rangle\rangle= 1$, then $\langle\langle \gamma|\hat d$ must be the bra for $|\gamma\rangle\rangle$. Therefore, each bosonic-fermionic pair of non-$\mathcal{Q}$-symmetric eigenstates
\begin{eqnarray}
\label{PairOfStates}
|\gamma\rangle\rangle,  \quad   \langle\langle \gamma|\hat d, \text{ and }
\hat d |\gamma\rangle\rangle, \quad \langle\langle \gamma|,
\end{eqnarray}
is defined by a single bra-ket pair, $\langle\langle \gamma|$ and $|\gamma\rangle\rangle$. Notably, $\langle\langle \gamma| \gamma\rangle\rangle = 0$ because the sum of the ghost degree's of the two forms (the bra and ket) is $D-1<D$. In result, the resolution of the unity operator from Eq.(\ref{Unity1}) is:
\begin{eqnarray}
\hat {\bm 1}_\mathcal{H} = \sum\nolimits_{\theta} |\theta \rangle\rangle \langle\langle \theta| + [\hat d, \sum\nolimits_{\gamma} |\gamma \rangle\rangle \langle\langle \gamma| ]_+.\label{unity}
\end{eqnarray}
We can now insert the resolution of unity of the Hilbert space at the temporal "infinity", $t=0,T$, of the path-integral in Eq.(\ref{pathintegral0}) and arrive at the famous expression for the Witten index:
\begin{eqnarray}
\mathcal{W} &=& \sum\nolimits_{\alpha}\int_{\Phi} \langle\langle \alpha| e^{\{\mathcal{Q}, \Theta\}}(-1)^{\hat F}|\alpha \rangle\rangle\nonumber\\
&=& \text{Tr }(-1)^{\hat F} e^{-T\hat H}.\label{Witten1}
\end{eqnarray}
Here, the pathintegrals in the first line are over the paths that connect the arguments of the bras and kets. In a certain sense, we broke the periodic boundary conditions for the ghosts and this is the reason for the emergence of the $(-1)^{\hat F}$ operator (see, \emph{e.g.}, Sec. 10.2.2. of Ref.[\onlinecite{MirrorSymmetry}]), where $\hat F = \chi^i\hat\partial_{\chi^i}$ is the ghost number operator, which is commutative with $\hat H$ so that the ghost number is a good quantum number: $\hat F|\alpha\rangle\rangle = F_{\alpha}|\alpha\rangle\rangle$, with $F_\alpha$ being an integer between $0$ and $D$.

The contribution into the Witten index from the non-$\mathcal Q$-symmetric bosonic-fermionic pairs, $\gamma's$, is zero. Only $\mathcal{Q}$-symmetric states, $\theta$'s, contribute:
\begin{eqnarray}
\mathcal{W} &=& N^+_\theta - N^-_\theta,
\end{eqnarray}
where $N_\theta^{\pm}$ are the numbers of bosonic/fermionic $\theta$'s.

\section{DS theory interpretation of the TFT objects}
\label{TFTTODS}
In the previous section, the approximation-free Parisi-Sourlas stochastic quantization procedure leading from a stochastic DS to its TFT representation was discussed. In this section, we will talk about the physical meanings of the TFT's objects from the DS theory point of view.

\subsection{Witten index and partition function of noise}
\label{WittenvsPartFunction}

Let us now get back to the definition of the Witten index in Eq.(\ref{WittenOriginal}). With the aid of the infinite-dimensional generalization of the formula of the integration of $\delta$-functions: $\int g(x)\delta(f(x))dx = \sum\nolimits_{x_\alpha,f(x_\alpha)=0}g(x_\alpha)/|f'(x_\alpha)|$, Eq.(\ref{windingnumber}) can be rewritten as:
\begin{eqnarray}
[w.n.] &=& \sum\nolimits_{\alpha} \text{sign }J(\varphi_\alpha) = N^+ - N^-,\label{widningnumber1}
\end{eqnarray}
where the summation is assumed over the (time-periodic) solutions of the SDE, $\varphi_\alpha$, at a given configuration of noise, \emph{i.e.}, $\varphi_\alpha$ are such that $F^i(t,\varphi_\alpha,\xi)=0$, and $N^\pm$ is the number of the solutions with the positive and negative Jacobians respectively.

Object $[w.n.]$ is of topological origin, that can be interpreted as an infinite-dimensional generalization of the Poinc\'are-Hoft theorem. The later states that for any continuous vector field on a compact finite-dimensional manifold without boundary, the sum of the indices (\emph{i.e.}, the sings of the determinants of the matrices of the second derivatives) over the critical points equals the Euler characteristic of that manifold. The manifold in our case is $\mathcal{M}_\varphi$, the vector field is $F^i(t)\delta/\delta\varphi^i(t)\in TM_{\varphi^i(t)}(\mathcal{M}_\varphi)$, and the critical points are the solutions of the SDE, $F^i(t) = 0$. In other words, as we vary the flow, the metric on $M$, or the configuration of the noise, the solutions of the SDE appear and disappear in pairs with positive and negative Jacobians, thus leaving $[w.n.]$ always the same. It would be natural now to expect that $[w.n.]$ must be equal to the Euler characteristic of $\mathcal{M}_\varphi$. The space $\mathcal{M}_\varphi$, however, is infinite-dimensional and its Euler characteristic may not be well-defined.

\begin{figure}[t] \includegraphics[width=6.0cm,height=4.0cm]{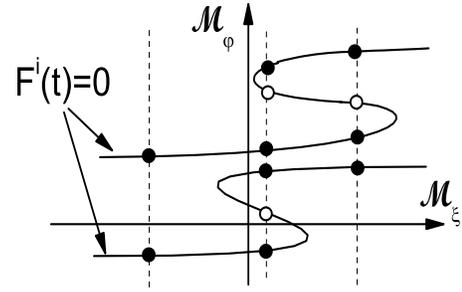}
\caption{\label{Figure1} Graphical demonstration of the topological nature of the index [$w.n.$] of the Nicolai map (Eq.(\ref{NicolaiMap})) defined by the SDE. Solutions come in pairs with positive (black dots) and negative (hollow dots) Jacobians, so that [$w.n.$] does not depend, \emph{e.g.}, on the configuration of the noise (dotted vertical lines), and in case given equals 2.}
\end{figure}

Nevertheless, the topological nature of $[w.n.]$ can be seen from yet another perspective. The point is that each $\varphi(t)\in\mathcal{M}_{\varphi}$ uniquely defines the configuration of the noise that makes this $\varphi^i(t)$ the solution of the SDE:
\begin{eqnarray}
\xi^a(\varphi(t)): \mathcal{M}_\varphi\to\mathcal{M}_\xi,\label{NicolaiMap}
\end{eqnarray}
where $\xi^a(\varphi(t))= e^a_i(\varphi(t)) ( \partial\varphi^i(t) + A^{i}(\varphi(t)) )$ from Eq.(\ref{SDE}). Eq.(\ref{NicolaiMap}) can be looked upon as a Nicolai map from $\mathcal{M}_\varphi$ to $\mathcal M_\xi$, and $[w.n.]$ equals to the winding number of the map, \emph{i.e.}, the number of times $\mathcal{M}_\varphi$ wraps around $\mathcal{M}_\xi$. As is clear from Fig.\ref{Figure1}, the winding number is the difference between the numbers of solution of the SDE with positive and negative Jacobians of the variable transformation, $\delta\xi^a(t)/\delta\varphi^i(t')$. The sign of this Jacobian is the same as the sign of that in Eq.(\ref{Jacobian}). To see this, one must think of $F^i(t)=0$ as of a constraint. The linearization of this constraint near the solution $F^i(t)=0$ gives:
\begin{eqnarray}
\int_{t'}\left(\frac{\delta F^i(t)}{\delta\varphi^j(t')}\delta \varphi^j(t') + \frac{\delta F^i(t)}{\delta\xi^a(t')}\delta\xi^a(t')\right) = 0.
\end{eqnarray}
Noting that $\delta F^i(t)/\delta\xi^a(t') = - e^i_a(\varphi(t))\delta(t-t')$, we get:
\begin{eqnarray}
\frac{\delta \xi^a(t)}{\delta\varphi^j(t')} = e^a_i(\varphi(t))\frac{\delta F^i(t)}{\delta\varphi^j(t')}   .
\end{eqnarray}
Because det $e^a_i(\varphi) = \sqrt{g}$, where $g=\det{g_{ij}}$, we have
\begin{eqnarray}
\text{det}\left(\frac{\delta \xi^a(t)}{\delta\varphi^j(t')}\right) = e^{(1/2)\int_{t=0}^T \text{log } g}\text{det}\left(\frac{\delta F^i(t)}{\delta\varphi^j(t')}\right).
\end{eqnarray}
The factor $e^{(1/2)\int_{t=0}^T \text{log } g}$ is always positive, so that the signs of the two sides of the previous equality are the same and consequently the winding number of the Nicolai map does equal Eq.(\ref{widningnumber1}).

Now, that the independence of $[w.n.]$ from the configuration of noise is established, we have
\begin{eqnarray}
\mathcal{W} = \langle\langle (N^+ - N^-) \rangle\rangle_\text{Ns} = [w.n.]\times \mathcal{Z}_\text{Ns},\label{WittenIndex}
\end{eqnarray}
where $\mathcal{Z}_\text{Ns}$ is from  Eq.(\ref{partitionfunctionofnoise}). Eq.(\ref{WittenIndex}) is the generalization of the claim in Ref.[\onlinecite{ChaosPaper}] that $\mathcal{W}$ equals the partition function of the noise. The later statement is exact for SDE's with $[w.n.]=1$, such as SDE's with Euclidian $M=\mathbb{R}^D$. An example of models with $[w.n.]\ne1$ is the SDE's with compact boundaryless $M$, for which $\mathcal{W}$ equals the Euler characteristic of $M$, as is known from the Witten models.

Its physical interpretation as of the partition function of the noise also suggests that $\mathcal W$ does not depend even on whether $\mathcal{Q}$-symmetry is broken or not. Indeed, the noise's partition function does not depend on the flow and the metric on $M$, and consequently on whether the ground states are $\mathcal Q$-symmetric or not.

\subsection{Partition functions of TFT and DS}
\label{PhysPart}
Any supersymmetric theory has two fundamental partition-function like objects: the Witten index and the physical partition function:
\begin{eqnarray}
\mathcal{Z} = \text{Tr } e^{-T\hat H}.\label{PartFuncZ}
\end{eqnarray}
The essence of $\mathcal Z$ can be deduced by comparing it with $\mathcal W$. As is seen from Eq.(\ref{WittenIndex}) and (\ref{Witten1}), the non-$\mathcal{Q}$-symmetric bosonic/fermionic pairs of states represent roughly the pairs of solutions of SDE with positive and negative $J$'s. If we, however, are interested in a stochastic expectation value, we have to treat all the solutions on equal footing disregard of the sign of their Jacobians. In other words, $\mathcal Z$ can be roughly interpreted as the stochastic expectation value of the total number of solutions of the SDE:
\begin{eqnarray}
\mathcal{Z} \sim \langle\langle (N^+ + N^-) \rangle\rangle_\text{Ns}.\label{PartFunc2}
\end{eqnarray}
We would like to emphasize that the strict equality sign is not appropriate in Eq (\ref{PartFunc2}). This can be seen in the simplest model - the 1D Langevin SDE with the harmonic potential with the interlevel separation $\omega_0$. For this model, $\mathcal{Z}=\text{cth}(T\omega_0/2)$, while $\langle\langle (N^+ + N^-) \rangle\rangle_\text{Ns}=1$, which follows from the fact that in this linear model any noise's configuration corresponds only to one solution of SDE. 

It is the trace of the evolution operator, \emph{i.e.}, $\mathcal{Z}$, which we believe must be viewed as the fundamental partition function of the DS.

In accordance, the normalized stochastic expectation value (SEV) is:
\begin{eqnarray}
\overline{\mathcal{O}(\Phi)} = \mathcal{Z}^{-1}\sum\nolimits_{\alpha}\int_{\Phi} \langle\langle \alpha| \mathcal{O}(\Phi) e^{\{\mathcal{Q}, \Theta\}}|\alpha \rangle\rangle,\label{Average2}
\end{eqnarray}
where $\mathcal O(\Phi)$ is an observable. The SEV can be obtained from what could be called the topological expectation value in Eq.(37) of Ref.[\onlinecite{ChaosPaper}] by complementing each observable with $(-1)^{\hat F}$. The role of operator $(-1)^{\hat F}$ can be interpreted as to unfold the partition function of the noise, \emph{i.e.}, the Witten index, into that of the DS.

Of physical importance is the limit of the infinitely long temporal evolution, $T\to\infty$. In this limit, only the ground states contribute into Eq.(\ref{Average2}), and the SEV becomes the vacuum expectation value (VEV):
\begin{eqnarray}
\overline{\mathcal{O}(\Phi)} = \mathcal{Z}^{-1} \sum\nolimits_{g} \int_\Phi \langle\langle g| \mathcal{O}(\Phi) e^{\{\mathcal Q,\Theta\}} |g\rangle\rangle,\label{average}
\end{eqnarray}
where $g$ runs only over the ground states, $\mathcal{Z}$ equals (up to an oscillating factor) the number of the ground states, and the pathintegral is assumed over the trajectories that connect the arguments of $|g\rangle\rangle$'s and $\langle\langle g|$'s at positive and negative temporal infinities.

To understand which eigenstates are the ground states, allow us to briefly recall the spectrum of the model discussed previously in Ref.[\onlinecite{ChaosPaper}]. From straightforward physical arguments, the real part of all eigenvalues, $\mathcal{E}_\alpha = \Gamma_\alpha + iE_\alpha$, is non-negative, $\Gamma_\alpha\ge0, \forall \alpha$. Eigenstates with $\Gamma_\alpha>0$ are dissipative and for this reason are not physical. These states will drop out from the partition function in the physical $T\to\infty$ limit. Only the non-dissipative states with $\Gamma_\alpha=0$ will survive the FP evolution and for this reason can be considered the physical states. Allow us to denote the physical part of the Hilbert space by $\mathcal H_p$.

The $\mathcal Q$-symmetric states, $\theta$'s, with zero-eigenvalues certainly belong to $\mathcal{H}_{p}$. If there are no other physical states in $\mathcal{H}_p$, then $\mathcal Q$-symmetry is not broken and all the $\theta$'s are the ground states of the model. If, however, $\mathcal{H}_{p}$ also contains physical non-dissipative Ruelle-Pollicott resonances (eigenstates with purely imaginary eigenvalues, $\mathcal{E}_\alpha^\pm = \pm iE_\alpha$) then $\mathcal Q$-symmetry is broken. In this case, the ground states are the bosonic-fermionic pair of the Ruelle-Pollicott resonance with the smallest (negative) "energy", $E_g=\min_{\alpha\in\mathcal{H}_p}E_\alpha$. This can be seen by Wick-rotating time "a little", $T\to T(1-i0^+)$, which will eliminate all the physical states from the partition function except for these ground states.

\subsection{Goldstone theorem and emergent long-range correlations}
\label{Goldstone}
Using Eq.(\ref{average}), we can now prove the Goldstone theorem that ensures that DS's with spontaneously broken $\mathcal Q$-symmetry must exhibit long-range correlations. For simplicity, we assume that there is no accidental degeneracy in the ground states so that there are only two ground states that according to Eq.(\ref{PairOfStates}) can be represented as $\langle\langle 0|\hat d$ and $|0\rangle\rangle$, and $\langle\langle 0|$ and $\hat d |0\rangle\rangle$. There exist a local $\mathcal Q$-exact observable, $\{\mathcal {Q}, X(t_0)\}$, (\emph{e.g.}, the Hamiltonian $H(t_0) = \{\mathcal{Q}, \bar Q(t_0)\}$) such that its VEV is non-zero:
\begin{eqnarray}
\overline{\{\mathcal{Q}, X(t_0)\}} = \int_\Phi \langle\langle 0| \{\mathcal{Q}, X(t_0)\} e^{\{\mathcal Q,\Theta\}} \hat d|0\rangle\rangle= c_X.\label{average1}
\end{eqnarray}
Here we used that $\mathcal{Z}=2$ and that we can move operator $\hat d$ through the $\mathcal{Q}$-exact expression between the bra and ket of the ground states, so that the contributions from the two ground states are the same. For the same reason, the expectation value of any $\mathcal Q$-exact operator within the Witten index is always zero, even when $\mathcal Q$-symmetry is broken. Therefore, the realization of the fact that it is the partition function of the TFT that must be used for the evaluation of VEV's is important, in particular, for the proof of the Goldstone theorem.

Let us now consider the following VEV:
\begin{eqnarray}
\overline{X(t_0)} = \int_\Phi \langle\langle 0| X(t_0) e^{\{\mathcal Q,\Theta\}} \hat d|0\rangle\rangle = 0,\label{average2}
\end{eqnarray}
which is of cause zero because $X(t)$ has ghost degree $(-1)$. One can make an infinitesimal $\mathcal Q$-rotation of the fields within the time domain $t'>t_0>t''$:
\begin{eqnarray}
\delta \Phi = \int_t \epsilon(t) \{\mathcal{Q}, \Phi(t)\},
\end{eqnarray}
where $\epsilon(t) = \epsilon \to 0$ if $t'>t_0>t''$ and $\epsilon(t) = 0$ otherwise. This changes the action and operator, $X$, as:
\begin{eqnarray}
\delta \{\mathcal Q,\Theta\} &=& \int_t (\partial_t\epsilon(t)) d(t) = \epsilon(-d(t') + d(t'')),\\
\delta X(t_0) &= & \epsilon \{\mathcal{Q}, X(t_0)\},
\end{eqnarray}
where $d(t)\equiv iB_i(t)\chi^i(t)$ is the N\"other charge of $\mathcal Q$-symmetry, that was introduced previously and identified as the exterior derivative. The VEV in Eq.(\ref{average2}) should not be changed because the $\mathcal Q$-rotation is only a change in the variables of integration. Leaving only the terms linear in $\epsilon\to0$ we arrive at
\begin{eqnarray}
\int_\Phi \langle\langle 0| X(t_0) (-d(t') + d(t'')) e^{\{\mathcal Q,\Theta\}} \hat d|0\rangle\rangle = c_X,
\end{eqnarray}
with $c_X\ne0$ from Eq.(\ref{average1}). One of the terms in the lhs, the one with $d(t'')$ acting on the $\hat d$-exact ket earlier than $X(t_0)$, vanishes because of the nilpotency of $\hat d$. Therefore,
\begin{eqnarray}
\overline{d(t') X(t_0)} = - c_X \ne 0.\label{GoldstoneFinale}
\end{eqnarray}
In other words, the correlation between $d(t')$ and $X(t_0)$ is infinitely long because $t'-t_0>0$ is arbitrary large.

We can as well chose $X(t_0)=i\epsilon' \bar\chi_i(t_0)\tilde A^i(t_0)$, with $\epsilon'$ being yet another small constant. This choice of $X(t_0)$ can be interpreted as a momentarily perturbation of the DS at $t_0$ with some additional flow field $A^i(t)\to A^i(t) + \epsilon'\delta(t-t_0)\tilde A^i(t_0)$. According to Eq.(\ref{GoldstoneFinale}), this perturbation will be correlated with $d(t')$ for infinitely long time. Operator $\hat d(t')$, in turn, has as its part the momentum operator, $\hat B_i(t')$, which is the quantum counterpart of the classical concept of velocity.

In the deterministic limit, this infinitely long, perturbation-induced "shift" in velocity will be gradually accumulated in the trajectory of the DS as time goes on. Even for $\epsilon'\to0$, at sufficiently large $t'-t_0$ the trajectory must become very different from the one that the DS would have developed if unperturbed. This is most likely the TFT's way of encoding the concept of high sensitivity to initial conditions in chaotic deterministic DS's.

In the general, stochastic case, the long-range correlations ensured by the Goldstone theorem mean that the DS is "complex". As it was discussed in Ref.[\onlinecite{ChaosPaper}], just as other supersymmetries, $\mathcal Q$-symmetry must be perturbatively stable. The other two possible levels of $\mathcal Q$-symmetry breaking, \emph{i.e.}, the mean-field (or rather Gaussian) and the (anti-)instanton condensation levels, must be interpreted as chaotic and intermittent complex dynamics respectively. Those are the two major sub-phases of complexity as is known from the Literature.

\subsection{TFT's wavefunctions and conditional probability densities}
\label{ConditionProbabilities}
For a $\psi^{(D)}$, the generalized FP Eq. becomes the conventional FP Eq.:
\begin{eqnarray}
\partial_t \psi^{(D)} = - \hat d (\hat j \psi^{(D)}),\label{conventionalFPEq}
\end{eqnarray}
because $\hat d\psi^{(D)}=0$ for any $\psi^{(D)}$ and $\hat j$ is from Eq.(\ref{current}). Eq.(\ref{conventionalFPEq}) has the meaning of the continuity equation and $\psi^{(D)}$ has the meaning of the total probability density. The expression in the parenthesis on the rhs must be interpreted as the current of the total probability density. The current is a form of $D-1$ degree that points onto the possibility that all wavefunctions of this degree may have the meaning of currents. \cite{PoincareDuals,ChaosPaper}

In fact, the only conclusion we can draw from the previous analogy with the continuity equation is that if a wavefunction of non-trivial ghost content, $\psi^{(n)}, n<D$, has the meaning of some physical quantity, then $\hat j \psi^{(n)} $ must have the meaning of the current of that physical quantity. In other words, $\hat j$ is the current operator as discussed, \emph{e.g.}, in Refs.[\onlinecite{PoincareDuals}] or [\onlinecite{Sinitsyn}].

We can use now the so "liberated" freedom in the interpretation of the wavefunctions, $\psi^{(n)},n<D$, and put all the wavefunctions on the same footing by claiming that the meaning of TFT's wavefunctions on the DS side is the conditional probability densities (CDP's). The total probability density, \emph{i.e.}, $\psi^{(D)}$, is a special case of CPD's. Before we will make an attempt to support this interpretation, however, we have to demonstrate that CPD's can be viewed as elements from the exterior algebra of $M$.

In the probability theory, a CPD is defined through a marginal probability density and a total probability density. For example:
\begin{eqnarray}
P_\text{cnd}^{(n)}(\varphi^1...\varphi^n|\varphi^{n+1}...\varphi^D)\times
\nonumber \\
\times P_\text{mrg}^{(D-n)}(\varphi^{n+1}...\varphi^{D}) = P_\text{tot}^{(D)}(\varphi).\label{CPD}
\end{eqnarray}
Here, $\varphi^{n+1}...\varphi^{D}$ are thought to be the already known variables, while $\varphi^1...\varphi^n$ are unknown.

The CPD in Eq.(\ref{CPD}) can be used, in particular, to evaluate the probability, $P^{(n)}_\text{cnd}(c_n)$, of finding the unknown variables within a n-chain, $c_n$, that belongs to a n-dimensional manifold formed in $M$ by fixing the known variables, $(\varphi^{n+1}...\varphi^D)\to \text{const}$:
\begin{eqnarray}\nonumber
P^{(n)}_\text{cnd}(c_n) = \int_{c_n} P^{(n)}_\text{cnd}(\varphi^1...\varphi^n|\varphi^{n+1}...\varphi^d)d\varphi^1\wedge...\wedge d\varphi^n.
\end{eqnarray}
There are $C^n_D=D!/(n!(D-n)!)$ CPD's of the $n^\text{th}$ "degree" and the same number of the corresponding differentials like $d\varphi^1\wedge...\wedge d\varphi^n$, depending on which $n$ out of $D$ variables we consider unknown.

We can formally consider the sum of combinations of all these CPD's with their corresponding differentials. This coordinate-free object is an element of the exterior algebra of $M$. An example of the coordinate-free representation of a CPD of the second degree in a 3-dimensional $M$ is this:
\begin{eqnarray}
P^{(2)}(\varphi) = P^{(2)}_{ij}(\varphi)d\varphi^i\wedge d\varphi^j,
\end{eqnarray}
where $P^{(2)}_{12}=-P^{(2)}_{21} = P(\varphi^1\varphi^2|\varphi^3)/2$, $P^{(2)}_{13}=-P^{(2)}_{31} = P(\varphi^1\varphi^3|\varphi^2)/2$, and
$P^{(2)}_{23}=-P^{(2)}_{32} = P(\varphi^2\varphi^3|\varphi^1)/2$.

The coordinate-free version of Eq.(\ref{CPD}) is:
\begin{eqnarray}
P^{(n)}_\text{cnd} \wedge P^{(D-n)}_\text{mrg} = P^{(D)}_\text{tot}.\label{DefCPD}
\end{eqnarray}
At this, the marginal probability density is a form that has no coordinate dependence in those directions, in which it is not a form (has no ghosts). This translates in the algebraic topology language into the following:
\begin{eqnarray}
\hat d P^{(n)}_\text{mrg} = 0.\label{MarginalCondition}
\end{eqnarray}
If $P^{(n)}_\text{cnd}$ also satisfies this condition, it is said that the variables of the CPD are independent of those of $P^{(D-n)}_\text{mrg}$, and $P^{(n)}_\text{cnd}$ itself can be looked upon as a marginal probability density.

In such situation, both the marginal and conditional probability densities must be non-trivial in the De Rahm cohomology of $M$ (must be $\hat d$-closed but not $\hat d$-exact), otherwise $\int_M P^{(n)}_\text{cnd}\wedge P^{(D-n)}_\text{mrg} = 0$ (for a compact $M$ with no boundary). In conventional notations of the algebraic topology, $[P^{(n)}_\text{cnd}]\in H^n$, $[P^{(D-n)}_\text{mrg}]\in H^{D-n}$, where $H$'s are the cohomologies of the corresponding degree. In this case, Eq.(\ref{DefCPD}) is a realization of the De Rahm cohomology ring. The interpretation of CPD's as of forms suggests that information theory and, \emph{e.g.}, intersection theory must be synergetic. This line of thinking could probably be called "information topology" by analogy with "information geometry". \cite{InformationGeometry}

We can now make an attempt to defend the idea that CPD's are the DS theory counterparts of the TFT's wavefunctions. One of the arguments comes from the shape of the kets of the $\mathcal Q$-symmetric states, $\theta$'s, in the deterministic limit. Those are the Poinc\'are duals of the unstable manifolds as was discussed briefly in the end of Sec. \ref{General} and, in more details, in Ref. [\onlinecite{ChaosPaper}]. If so, the unstable manifolds can be thought to parameterize the "known" variables, while the "stable" directions, in which the kets have ghosts, are the delta-function-like CPD's just as it should be.

There is yet another argument that points on the CPD's interpretation of the TFT's wavefunctions. This argument applies also for to stochastic DS's. The argument is the similarity between the definition of the CPD in Eqs.(\ref{MarginalCondition}) and Eq.(\ref{DefCPD}), and the Hilbert space structure discussed in Sec. \ref{General}. From Eqs.(\ref{unity}) and/or (\ref{PairOfStates}), it is seen that for any non-$\mathcal Q$-symmetric eigenstate either bra or ket satisfy Eq.(\ref{MarginalCondition}). The bras and kets, in turn, are "formed" by backward and forward temporal evolution and for this reason contain unstable and stable variables respectively. This means that the bras and kets have the meaning of pairs of marginal and conditional probability densities for the unstable and stable variables. For each non-$\mathcal Q$-symmetric state either bra or ket does not satisfy Eq.(\ref{MarginalCondition}), suggesting that the unstable and stable variables are not dynamically independent. In contrary, for a $\mathcal Q$-symmetric state, $\theta$, both bra and ket satisfy Eq.(\ref{MarginalCondition}). It can be said that the unstable and stable variables are dynamically independent.

From the above discussion the following interpretation of the spontaneous $\mathcal Q$-symmetry breaking follows - for the DS's with broken $\mathcal Q$-symmetry, the unstable and stable variables are not dynamically independent in the ground state(s).

\subsection{$\mathcal Q$-symmetry and Stokes' theorem}
\label{PhysQSym}
As a closing remark, allow us to repeat the argument from Ref.[\onlinecite{ChaosPaper}], that the $\mathcal Q$-symmetry is actually a necessity for the "extended" description of a DS, in terms of not only the total probability densities but also in terms of CPD's. Consider the Stokes' theorem:
\begin{eqnarray}
\int_{c_n} \hat d\psi^{(n-1)} = \int_{\partial c_n} \psi^{(n-1)},
\end{eqnarray}
where $c_n$ is some n-chain with boundary $\partial c_n$. This expression must hold for all times, while both $\hat d\psi^{(n-1)}(t) = e^{-t\hat H}\hat d\psi^{(n-1)}$ and $\psi^{(n-1)}(t)=e^{-t\hat H}\psi^{(n-1)}$ evolve according to the generalized FP equation. Therefore
\begin{eqnarray*}
\int_{c_n} e^{-t\hat H}\hat d\psi^{(n-1)} = \int_{\partial c_n} e^{-t\hat H} \psi^{(n-1)} = \int_{c_n} \hat d e^{-t\hat H}\psi^{(n-1)},
\end{eqnarray*}
which shows that the generalized FP Hamiltonian is commutative with $\hat d$. On the other hand, we known that $\hat d$ is the N\"other charge of $\mathcal Q$-symmetry. In other words, the theory with the extended description must possess $\mathcal Q$-symmetry, \emph{i.e.}, it must be a TFT.

\section{Conclusion}
\label{Conclusion}
In this paper, which is a continuation of the study of the relation between Witten-type topological field theories and dynamical systems theory, \cite{ChaosPaper} it was shown that on the DS theory side, the Witten index represents up a topological factor the partition function of the stochastic noise, while the physical partition function of the TFT is that of the DS. Also, it was shown that the wavefunctions of TFT must be interpreted as the conditional probability densities of the corresponding DS.

To our opinion, the story of the relation between TFT's and the DS's is reasonably reliable at this stage. At the same time, there are many questions that still need to be answered. One of them is the interpretation of observations/measurements. In other words, what does an observation do to a wavefunction? Does it automatically project the wavefunction onto the corresponding conditional probability density sector (ghost sector) because we learned values of some variables? If yes, how far is this projection from the concept of wavefunction collapse in quantum mechanics or from the Bayesian update of probability density? Other important questions are how to identify $\mathcal Q$-order parameters, what is their physical meaning, and what is their low-energy effective actions. Answering these questions will allow for the construction of effective theories of various complex DS's - probably one of the most important possible applications of the TFT of DS.

\acknowledgments

We would like to thank Hao-Yuan Chang, Robert N. Schwartz, and Kang L. Wang for discussions. The work was partly supported by Defense Advanced Research Projects Agency, Defense Sciences Office, Program: Physical Intelligence, contract HR0011-01-1-0008.


\begin{references}
\bibitem{ParisiSourlas} G. Parisi and N. Sourlas, Phys. Rev. Lett. {\bf 43}, 744 (1979).
\bibitem{Witten} E. Witten, Commun. Math. Phys. {\bf 117}, 353 (1988); {\bf 118}, 411 (1988).
\bibitem{TFT} D. Birmingham, M Blau, M. Rakowski, and G. Thompson, Phys. Reps. {\bf 209}, 129 (1991).
\bibitem{Frenkel} E. Frenkel, A. Losev, and N. Nekrasov, J. of Inst. of Math. Jussieu {\bf 10}, 463 (2011).
\bibitem{Labastida} J. M. F. Labastida, Commun. Math. Phys. {\bf 123}, 641 (1989).
\bibitem{ChaosPaper} I. V. Ovchinnikov, Chaos {\bf 22}, 033134 (2012).
\bibitem{Mostafozadeh} A. Mostafazadeh, J. Math. Phys. {\bf 43}, 2814 (2002).
\bibitem{FormsAndWavefunctions} E. Witten, J. Diff. Geom. {\bf 17}, 661 (1982).
\bibitem{Stone} M. Stone, \emph{Mathematics for Physics II}, (PIMANDER-CASAUBON Alexandria, Florence, London, 2001)
\bibitem{PoincareDuals} S. Tanase-Nicola and J. Kurchan, J. Stat. Phys. {\bf 116}, 1201 (2004).
\bibitem{MirrorSymmetry} K. Hori, S. Katz, A. Klemm, R. Pandharipande, R. Thomas, and R. Vakil, \emph{Mirror symmetry} (Clay mathematics monographs) (American Mathematical Society, 2003), v.1.
\bibitem{Sinitsyn} N. A. Sinitsyn, A. Akimov, and V. Y. Chernyak, Phys. Rev. E {\bf 83}, 021107 (2011).
\bibitem{InformationGeometry} S. Amari and H. Nagaoka, "Methods of information geometry", Translations of mathematical monographs; v. 191, American Mathematical Society, 2000 (ISBN 978-0821805312).
\end{references}
\end{document}